\documentclass[twocolumn, mathlines]{aastex7}
\usepackage{gensymb}
\usepackage{amsmath}

\DeclareRobustCommand{\ion}[2]{%
\relax\ifmmode
\ifx\testbx\f@series
{\mathbf{#1\,\mathsc{#2}}}\else
{\mathrm{#1\,\mathsc{#2}}}\fi
\else\textup{#1\,{\mdseries\textsc{#2}}}%
\fi}

\begin{document}

\title{Search for Faint Lone Double-Peaked H$\alpha$ Lines as IMBH Signatures in the MUSE Deep Field}

\author[0000-0002-5641-8102]{Jyoti yadav}
\affiliation{Instituto de Astrofísica de Canarias, Vía Láctea s/n, E-38205 La
Laguna, Spain}
\affiliation{Departamento de Astrofísica, Universidad de La Laguna, E-38206
La Laguna, Spain}
\email[show]{yadavjyoti636@gmail.com} 

\author[orcid=0000-0003-1123-6003,gname=Jorge, sname='  Jorge Sánchez Almeida ']{Jorge Sánchez Almeida}
\email{INSERT_EMAIL_HERE} 
\affiliation{Instituto de Astrofísica de Canarias, Vía Láctea s/n, E-38205 La
Laguna, Spain}
\affiliation{Departamento de Astrofísica, Universidad de La Laguna, E-38206
La Laguna, Spain}
\email[]{} 
\author[orcid=0000-0001-8876-4563,gname=Casiana, sname='  Casiana Muñoz Tuñón']{Casiana Muñoz Tuñón} 
\affiliation{Instituto de Astrofísica de Canarias, Vía Láctea s/n, E-38205 La
Laguna, Spain}
\affiliation{Departamento de Astrofísica, Universidad de La Laguna, E-38206
La Laguna, Spain}
\email[]{} 
\author[orcid=0000-0003-1803-6899,gname=Joao, sname='  Joao Calhau']{João Calhau} 
\affiliation{INAF-Osservatorio Astronomico di Capodimonte, Via Moiariello 16, 80131 Napoli, Italy}
\email[]{}

\begin{abstract}
Double-peaked H$\alpha$ emission profiles can serve as potential signatures of accreting intermediate-mass black holes (IMBHs), particularly those residing outside galactic nuclei. Such features are expected to arise from rotating disk-like structures around black holes and can be used to identify elusive IMBH candidates. \citet{Almeida2022ApJ...934..100S} reported a sample of such double-peaked H$\alpha$ sources in the MUSE-Wide survey, interpreting them as potential signatures of wandering IMBHs after systematically excluding alternative explanations. Their method relied on constructing H$\alpha$ maps around central galaxies and visually identifying compact emission clumps in the surrounding halo regions. In this work, we revisit the analysis using the deeper MUSE Extremely Deep Field (MXDF) data and an automated detection algorithm tailored to identify such features. However, we do not recover any candidate population in MXDF, resulting in a null detection. This outcome is nevertheless informative, as it (1) highlights the inherent challenges in detecting IMBHs, and (2) demonstrates the potential of automated approaches for future systematic searches, even though it did not yield a positive outcome in this case.
\end{abstract}


\section{Introduction}\label{sec:Introduction}
Over the past two decades, substantial progress has been made in understanding supermassive black holes (SMBHs) and their central role in galaxy evolution. Although SMBHs are minuscule compared to their host galaxies, gas accretion can make them observable as active galactic nuclei (AGN), often outshining the entire galaxy. The discovery of luminous quasars at redshifts $z>6$, hosting black holes with inferred masses of $10^{8}$–$10^{9}$ M${\odot}$ \citep{Mortlock2011Natur.474..616M, Banados2018ApJ...861L..14B, Inayoshi2020ARA&A..58...27I, Wang2021ApJ...907L...1W, Pacucci2022MNRAS.509.1885P}, has motivated the hypothesis that SMBHs originate from an earlier population of seed black holes formed at very high redshifts ($z \gtrsim 15$–20). These seeds are expected to span intermediate masses, ranging from $\sim$100 M${\odot}$, potentially originating from Population III stellar remnants \citep{Madau2001ApJ...551L..27M, Hirano2014ApJ...781...60H}, to $10^{4}$–$10^{5}$ M$_{\odot}$, possibly formed through direct gas collapse or dynamical processes in dense stellar environments \citep{Bromm2003Natur.425..812B, Inayoshi2020ARA&A..58...27I, Volonteri2021NatRP...3..732V}. These progenitors are commonly referred to as intermediate-mass black holes (IMBHs), which may subsequently grow via gas accretion and stellar interactions \citep{Rosswog2009ApJ...695..404R, Sakurai2017MNRAS.472.1677S}. Another way to form IMBHs is via runaway collisions of stars in dense stellar clusters \citep{Portegies2004Natur.428..724P, Devecchi2009ApJ...694..302D, Mapelli2016MNRAS.459.3432M}. This can produce BH masses of 10$^{3}$--10$^{4}$ M${\odot}$. Gravothermal collapse in a self-interacting dark matter halo can also lead to BH formation when the core becomes relativistic and dynamically unstable \citep{Balberg2002PhRvL..88j1301B, Feng2021ApJ...914L..26F, Roberts2025JCAP...01..060G}. For recent reviews on IMBHs, see, e.g., \citet{Mezcua2017IJMPD..2630021M, Greene2020ARA&A..58..257G, Volonteri2021NatRP...3..732V}.

In the cosmological context, seed black holes often re-
side within dark matter halos that undergo mergers over
time \citep{Matteo2023MNRAS.525.1479D}. After the halos merge, the
embedded black holes are drawn toward the center of the
merged system through dynamical friction \citep{Dong2022ApJ...929..120D}. However, in minor mergers the satellite halo is often tidally disrupted before coalescence, leaving its central black hole on a wide orbit. As a result, a population of wandering black holes is predicted to exist \citep{Tremmel2018MNRAS.475.4967T, Ricarte2021ApJ...916L..18R, Ricarte2021MNRAS.503.6098R}. Hyperluminous X-ray sources \citep{Barrows2019ApJ...882..181B, Matteo2023MNRAS.525.1479D} and spatially offset AGN \citep{Mezcua2020ApJ...898L..30M, Reines2020ApJ...888...36R} represent possible, though observationally challenging, signatures of this population.

Understanding the formation and early growth of SMBHs is therefore closely tied to identifying and characterizing IMBHs. Despite their theoretical importance, IMBHs remain observationally elusive with no confirmed detections to date. This could be due to their intrinsically faint emission and the difficulty of measuring dynamical signatures, especially in low-mass or distant systems. As a result, their formation channels, number densities, and physical properties remain poorly constrained. \citet{Greene2020ARA&A..58..257G} estimated an IMBH number density of $\sim$0.31 Mpc$^{-3}$ from globular clusters and $\sim$0.06–0.3 Mpc$^{-3}$ from disrupted satellites. Cosmological simulations predict substantial IMBH mass densities; for example, the Astrid simulation yields an integrated IMBH mass density of $\sim1.5\times10^{5}$ M${\odot}$ Mpc$^{-3}$ at $z\sim2$ for $M_{\rm BH}<10^{6}$ M$_{\odot}$ \citep{Matteo2023MNRAS.525.1479D}. Direct detection of this population at high redshift remains challenging, as these objects are not expected to be sufficiently luminous \citep{Pacucci2015MNRAS.454.3771P, Natarajan2017ApJ...838..117N}.

A wide range of observational efforts has therefore been pursued to search for IMBHs. Dynamical methods have employed stellar and gas kinematics \citep{Kormendy2013ARA&A..51..511K, Rix1997ApJ...488..702R, Gebhardt2003ApJ...583...92G}, proper motions \citep{McLaughlin2006ApJS..166..249M, MacLeod2016ApJ...819....3M, Mann2019ApJ...875....1M}, and hypervelocity stars \citep{Hills1988Natur.331..687H, Yu2003ApJ...599.1129Y, Zhang2013ApJ...768..153Z, Brown2018ApJ...866...39B, Koposov2020MNRAS.491.2465K}, targeting globular clusters \citep{Noyola2010ApJ...719L..60N, Jalali2012A&A...538A..19J, Baumgardt2017MNRAS.464.2174B} and galactic nuclei \citep{Nguyen2018ApJ...858..118N, Boyce2017ApJ...846...14B}. These efforts are complemented by searches for optical transients associated with accreting IMBHs \citep{Filho2018MNRAS.478.2541F}.

In the Milky Way, searches have focused on IMBH remnants from past dwarf galaxy mergers \citep{Maillard2004A&A...423..155M, Rashkov2014ApJ...780..187R}, while in external galaxies, accreting IMBHs have been studied in low-mass nuclei using optical spectroscopy and reverberation mapping \citep{Filippenko2003ApJ...588L..13F, Barth2004ApJ...607...90B, Greene2004ApJ...610..722G, Reines2013ApJ...775..116R, Peterson2005ApJ...632..799P, Woo2019NatAs...3..755W, Pucha2025ApJ...982...10P, Liu2025arXiv250312898L}. Off-nuclear IMBH candidates have also been explored through ultraluminous X-ray sources and globular cluster dynamics \citep{Ebisuzaki2001ApJ...562L..19E, Maccarone2005MNRAS.356L..17M, Maccarone2008MNRAS.389..379M, Strader2012ApJ...750L..27S}, while wandering IMBHs may additionally be detected via tidal disruption events or gravitational-wave signals \citep{Wevers2017MNRAS.471.1694W, Wevers2019MNRAS.487.4136W}.

When actively accreting, IMBHs may exhibit isolated double-peaked H$\alpha$ emission profiles, often associated with rotating or kinematically disturbed accretion disks \citep{Smak1969AcA....19..155S}. Similar features are observed in X-ray binaries and cataclysmic variables \citep{Grundstrom2007ApJ...660.1398G, Zolotukhin2011A&A...526A..84Z}, as well as in the broad-line regions of AGN \citep{Eracleous1994ApJS...90....1E, Ho2008ARA&A..46..475H}. Recent JWST observations have revealed compact high-redshift sources (“little red dots”) whose H$\alpha$ spectra show comparable signatures \citep{Matthee2024ApJ...963..129M, Kocevski2025ApJ...986..126K}. Building on this framework, \citet{Almeida2022ApJ...934..100S} developed a new approach complementary to previous studies that focussed on detecting isolated double-peaked H$\alpha$ emission in the MUSE-Wide survey and identified 118 potential wandering IMBH candidates.

In the present work, we extend the approach of \citet{Almeida2022ApJ...934..100S} to the significantly deeper MUSE Extremely Deep Field (MXDF), providing a stringent evaluation of the double-peaked H$\alpha$ method. This enables us to assess its robustness under improved sensitivity and to constrain the rates of false-positive and false-negative detections which is a critical factor in blind IMBH searches. We focus on double-peaked H$\alpha$ emission up to a wavelength of 8950\,\AA, corresponding to $z\approx0.35$, targeting accreting IMBHs that may reside outside galactic nuclei, such as remnants of disrupted satellites, black holes dynamically ejected from their hosts, or primordial seeds wandering within their original dark matter halos. We place particular emphasis on identifying and controlling spurious detections, a key limitation in blind searches for faint IMBH signatures.

H$\alpha$ is one of the strongest and most reliably detected optical emission lines within the MUSE spectral range, making it especially well suited for probing faint and compact sources. Its relatively high signal-to-noise ratio enables the identification of subtle kinematic structures such as double-peaked profiles that may arise from gas in the vicinity of accreting black holes. While other emission lines, including [\ion{O}{iii}]~$\lambda5007$ or H$\beta$, can also exhibit double-peaked morphologies \citep{Fu2012ApJ...745...67F, Qiu2025ApJ...991...14Q, Pucha2025ApJ...982...10P}, these lines are often significantly fainter in low-luminosity, off-nuclear systems, making H$\alpha$ the most robust tracer for identifying such IMBH candidates in deep MUSE observations. Our work complements previous studies by assessing the robustness of the double-peak detection method and clarifying the limitations of double-peaked H$\alpha$ emission as a tracer of wandering IMBHs.

We develop and apply a carefully tuned automated algorithm optimized to detect double-peaked profiles with wavelength separations below 20\,\AA. Given the faintness and complexity of such signals, the method is designed to minimize false positives, highlighting the intrinsic challenges of identifying IMBHs through weak emission-line signatures.

The paper is organized as follows. In Sect.~\ref{sec:data}, we describe the MUSE datasets used in this study. Sect.~\ref{sec:method} presents the detection technique developed to identify faint double-peaked emission in noisy spectra. Sect.~\ref{sec:application} discusses the application of the algorithm to the MUSE data. In Sect.~\ref{sec:comparison}, we compare our results with previous studies, followed by discussion and conclusions in Sect.~\ref{sec:discussion} and Sect.~\ref{sec:conclusion}.

\section{Data}\label{sec:data}

MUSE \citep{Bacon2014Msngr.157...13B} is a second-generation instrument mounted on the Very Large Telescope. MUSE is an  optical integral field unit (IFU) which performs integral field spectroscopy over 4750–9350\,\AA\ wavelength range. In wide field mode, it offers a 1$\arcmin$ $\times$ 1$\arcmin$ field of view with a spatial sampling of 0.2$\arcsec$. MUSE has a spectral sampling of 1.25\,\AA\ and each exposure yields nearly 90,000 individual spectra. MUSE provides deep, spatially resolved, and high-quality IFU data over a wide field of view, making it uniquely suited for detecting faint and spatially offset emission-line sources such as accreting IMBH. The combination of spectral and spatial information allows us to identify double-peaked emission features, determine their spatial origin, and distinguish genuine astrophysical signals from artifacts or noise. We use the archival dataset from three complementary MUSE surveys -- MXDF, MOSAIC, and MUSE-Wide to take advantage of their different depths and areas.

We utilized archival MUSE data from the MXDF survey \citep{Bacon2023A&A...670A...4B}, which is one of the deepest spectroscopic observations ever conducted in a 8-m class telescope, with a total exposure time of 141 hours. We use the MXDF dataset to search for IMBH candidates by identifying double-peaked H$\alpha$ emission lines. The MXDF is centered on the Hubble Ultra Deep Field region, specifically targeting the area with deep WFC3 imaging \citep{Illingworth2013ApJS..209....6I}. Observations were carried out with VLT/MUSE equipped with the adaptive optics facility and GALACSI, its ground-layer adaptive optics module \citep{Kolb2016SPIE.9909E..2SK, Leibundgut2017Msngr.170...20L, Madec2018SPIE10703E..02M}. The central coordinates of the field are RA = 53.16467$^{\circ}$, Dec = -27.78537$^{\circ}$ (J2000, FK5). The resulting field of view is approximately circular in shape with a radius of 41\arcsec\ and 31\arcsec\ for respectively 10+ and 100+ h of depth. The FWHM of the MXDF data at 4850, 7000 and 9350\,\AA\ is 0.72\arcsec , 0.55\arcsec , and 0.47\arcsec\ respectively. The MUSE-MXDF data used in this study was obtained from the ESO archive, as provided by Roland Bacon (Programme ID: 1101.A-0127), with a total exposure time of 502500 seconds. 

We used the source catalogue provided by \citet{Bacon2023A&A...670A...4B}, which lists all objects detected within the MUSE-MXDF field. This catalogue was employed to cross-match and exclude detections associated with known sources such as galaxies, stars, or other identified objects from our sample.

\citet{Almeida2022ApJ...934..100S} identified 118 double-peaked H$\alpha$ emission-line candidates in the MUSE-Wide survey, which were proposed as potential signatures of IMBHs. In this work, we use archival MUSE-Wide dataset \citep{Urrutia2019A&A...624A.141U}, along with the IMBH candidate sample from \citet{Almeida2022ApJ...934..100S}, to investigate the properties of these sources in greater detail. MUSE-Wide is a relatively shallow survey conducted with MUSE in the CANDELS/GOODS-S and CANDELS/COSMOS regions. The final survey will cover a total area of 100 $\times$ 1 arcmin$^{2}$ on the sky. The survey complements the existing MUSE-Deep observations in the Hubble Deep Field South \citep{Bacon2017A&A...608A...1B}. Each pointing in MUSE-Wide has an integration time of 1 hour. The current data release (DR1) includes 44 contiguous fields, each covering 1$\arcmin$ $\times$ 1$\arcmin$.

The MUSE-MOSAIC dataset partially overlaps with the MUSE-Wide survey, providing an opportunity to test whether the double-peaked H$\alpha$ detections reported by \citet{Almeida2022ApJ...934..100S} are also present in deeper observations. To perform this comparison, we used archival MUSE-MOSAIC data, a medium-deep field with an exposure time of approximately 10 hours \citep{Inami2017A&A...608A...2I}. The observations cover a 3$\times$3 arcmin${^2}$ area. This field is composed of a mosaic of nine individual MUSE pointings (udf-01 to udf-09), each with a 10-hour integration.

\section{Method}\label{sec:method}
We employ a single Gaussian kernel designed to suppress noise and enhance the detection of double peaks with the expected separation. In the first step, we test the kernel on synthetic spectra to optimize the parameters for reliable detection. Finally, we apply the kernel with the optimized parameters to the MUSE-MXDF dataset through cross-correlation in order to search for double-peaked signals.

\subsection{Description of kernel}\label{sec:simple_kernel}
To reliably identify double-peaked emission lines with the right separation while filtering out single-peaked profiles, we adopted a simplified method based on cross-correlating the spectra with a single Gaussian kernel. When the kernel aligns with an emission feature, it produces a strong positive correlation peak, whereas regions without lines show weak correlation. This approach enhances the detection of weak emission-line features while dumping the random noise. We systematically explored and optimized the kernel parameters to minimize false detections. The gaussian kernel is defined as follows:

\begin{equation}
    g(x)= \frac{A}{\sqrt{2\pi} \sigma_{gauss}}  \times e^{-\frac{1}{2}(\frac{x-\mu}{\sigma_{gauss}})^2}
\end{equation}
$x$ is the  wavelength. We adopted $\mu$ = 6562.8\,\AA\ for the rest-frame wavelength of H$\alpha$ and $\sigma_{gauss}$ is defined as follows:
\begin{equation}
    \sigma_{\textrm{gauss}} = \frac{\textrm{FWHM}}{2\sqrt{\textrm{2ln(2)}}}
\end{equation}

The FWHM is the full width at half maximum of the kernel. To determine the optimal values of the free parameters of the kernel for best signal identification, we generated artificial test spectra with added random gaussian noise. The artificial spectra were constructed to closely reproduce the characteristics of real MUSE data. We first defined a linear continuum with a small slope over a wavelength grid similar to MUSE spectrum, thereby preserving the same wavelength range and sampling. Random Gaussian noise was then added using NumPy’s random number generator. We added gaussian profiles with amplitudes defined as multiples of the standard deviation of the underlying noisy signal. This approach allows us to control the height of artificial emission peaks relative to the background noise, effectively setting an amplitude-to-noise ratio. We call this S/N ratio from now on. As an initial test, we used an amplitude equal to four times the noise standard deviation, and later varied this factor to assess the robustness of the process. With the selection procedure defined above, we evaluated its ability to reliably detect double peaks in spectra and examined how this performance depends on kernel parameters.

We tested the algorithm for several kernel free parameters for e.g. gaussian width, the correlation cutoff threshold, and the S/N ratio of the emission lines in the test spectrum. We varied the parameters to maximize the effectiveness of the algorithm. We discuss about maximizing the effectiveness of the kernel in section~\ref{sec:effectiveness_simple_kernel}.

The algorithm works in two main steps as follows. First we cross-correlated the synthetic spectra with the simple kernel using \texttt{scipy.signal.correlate} function to evaluate its effectiveness in detecting double peaks. After computing the correlation function, we calculated its mean and standard deviation, then apply a 1$\sigma$ cutoff. We recalculated the mean and standard deviation of 1$\sigma$ clipped correlation function. These refined values define the final threshold (x $\sigma$) above which a signal is considered a valid peak candidate. 
Second we identified the peaks exceeding this threshold in the correlation spectrum using \texttt{scipy.signal.find$\_$peaks}. The single gaussian kernel produces positive correlation peaks for both single and double emission lines. We identify a double-peaked profile when two positive correlation peaks are found within 20\,\AA\ of each other around the expected emission position. If only one peak falls within this range and the second lies beyond 20\,\AA\, the detection is classified as a single-peaked line. The 20\,\AA\  is chosen as it corresponds to velocity separations well within the resolving power of MUSE, ensuring sensitivity to genuine double-peaked features while reducing the likelihood of misclassifying widely separated or unrelated peaks. It is also the separation ideal to detect the signals assigned to IMBH in \citet{Almeida2022ApJ...934..100S}. We note that the physical velocity separation of double-peaked lines can vary with the system properties, and the fixed 20\,\AA\ is a practical compromise for a blind search across the redshift range of interest. Adopting a fully redshift-dependent separation would complicate the automated detection without significantly improving sensitivity for the faint, isolated H$\alpha$ lines we target.

\begin{figure*}
    \centering
    \includegraphics[width=0.98\linewidth]{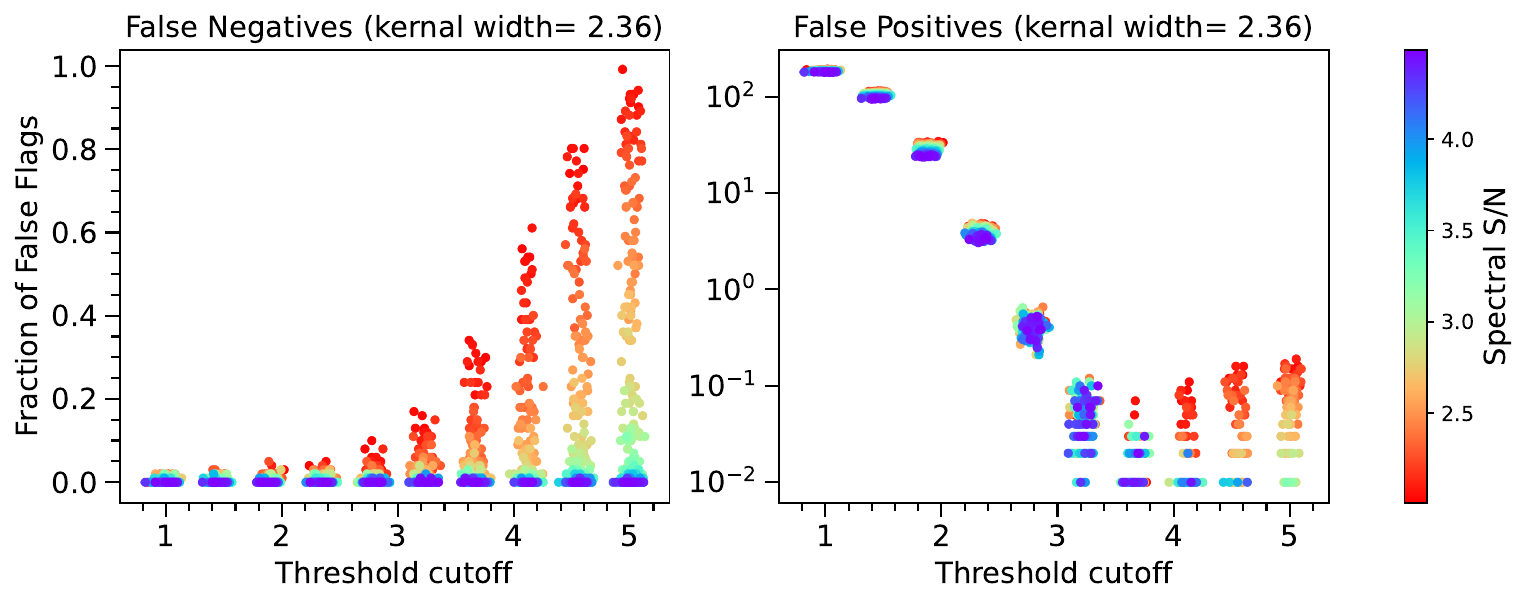}\\
    \includegraphics[width=0.98\linewidth]{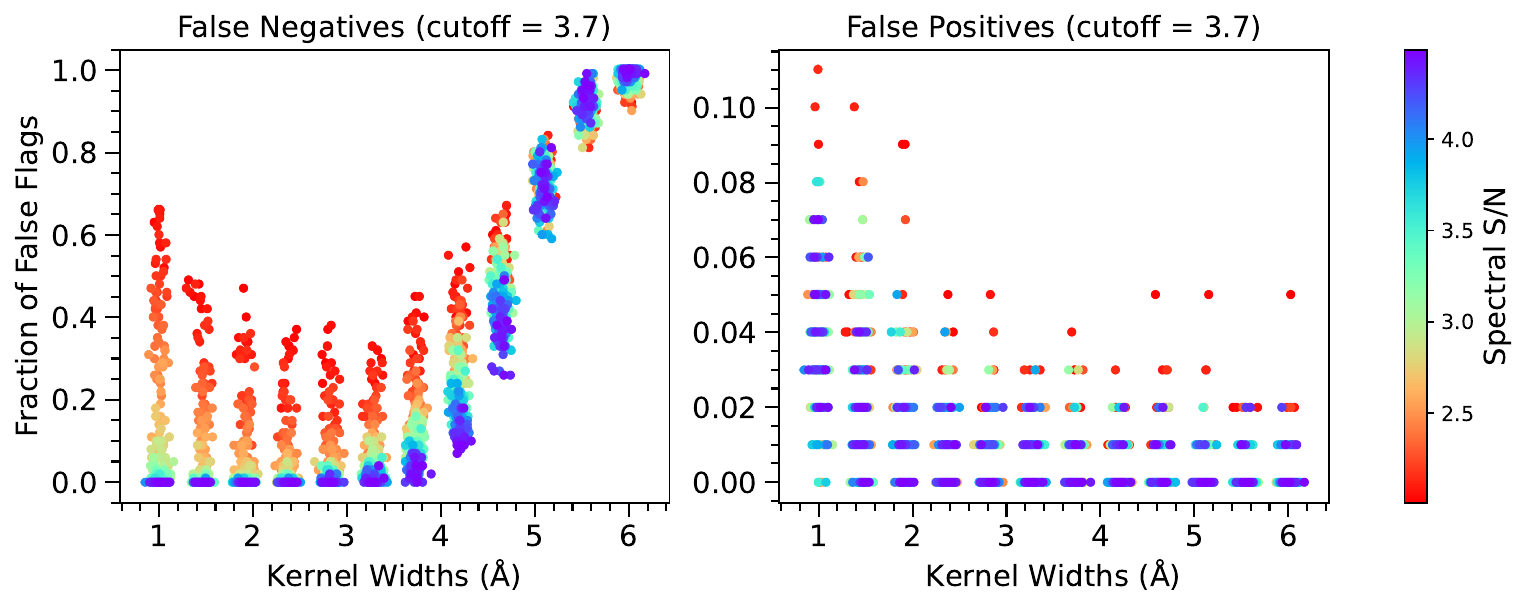}
    \caption{Top panels show the false negative (left) and false positive (right) flags as a function of threshold cutoff, with the gaussian kernel width ($\sigma_{gauss}$) fixed at 2.36\,\AA\ (the minimum allowed by spectral resolution of MUSE). These flags are generated during testing of the algorithm at varying S/N ratios in the artificial spectrum. Bottom panels shows the false negative (left) and false positive (right) flags as a function of kernel width while fixing the cutoff at 3.7$\sigma$ at varying S/N ratios in the artificial spectrum. The fraction of false flags represents the number of false positives or negatives per Monte Carlo run. Since multiple points overlap, we introduce a random jitter with standard deviation of 0.05 in the x axis.}
    \label{fig:single_kernel_test}
\end{figure*}

\subsection{Maximizing the effectiveness of kernel}\label{sec:effectiveness_simple_kernel}
We varied the free parameters mentioned in section~\ref{sec:simple_kernel} and estimated the number of false detections for each configuration, across a range of S/N ratios in the artificial spectra. This is crucial, as a high false-positive/negative rate in a blind search can lead to unreliable detections. To estimate these detection rates, we ran hundreds of correlation attempts on the artificial spectra for different values of the quantities while varying one parameter at a time, keeping others fixed. The S/N of the spectrum was varied between 0 and 20. We consider detections located outside ±20\,\AA\ of the expected wavelengths of the two components as false positives, while the absence of detections within this range is classified as a false negative.

Figure~\ref{fig:single_kernel_test} shows the fraction of false flags (number of false positives or negatives) per Monte Carlo run. A false negative can only occur once, as it corresponds to a single missed detection at the true location. In contrast, false positives can occur multiple times within the same spectrum, as random noise or artifacts may be incorrectly identified as double-peaked features. To find the optimum value of the cutoff limit, we fixed the gaussian width at approximately 2.4\,\AA\, which is the spectral resolution of MUSE. When finding the optimum value of gaussian width, we fixed the threshold cutoff at 3.7$\sigma$. This cutoff corresponds to the average S/N ratio expected from the emission distribution of the candidates identified by \citet{Almeida2022ApJ...934..100S}. The results are summarized in Figure~\ref{fig:single_kernel_test}. The top panel shows the false flags as a function of the threshold cutoff value, with the gaussian width fixed at 2.36\,\AA\, while the bottom panel shows the false flags as a function of gaussian kernel width, with the cutoff fixed at 3.7$\sigma$. The color bar shows the range of S/N ratios considered in the artificial spectra. We find that the false positive detection increases sharply when the cutoff is reduced below 3.0$\sigma$ (Figure~\ref{fig:single_kernel_test} top right panel). Specifically, at a cutoff of 2.8$\sigma$, approximately 40 out of every 100 runs result in a false positive detection, which is an unacceptably high rate.  In contrast, the false negative rates increases at higher cutoff values (Figure~\ref{fig:single_kernel_test} top left panel). For detection thresholds exceeding 4$\sigma$, the false-negative rate increases. This highlights a clear trade-off between minimizing false positives and false negatives. To prevent unreliable false positives, we decided to use a slightly higher cutoff threshold value for the detection of double peaks on MUSE-MXDF dataset. While this threshold may result in the loss of some lower-significance signals, our priority is to avoid false positives and thus we accept the potential loss of some lower-significance signals. We find that the optimal correlation cutoff lies between 3$\sigma$ and 4$\sigma$, which is consistent with the expected S/N distribution of sources in \citet{Almeida2022ApJ...934..100S}.

We also estimated the false detection rate as a function of the gaussian width (Figure~\ref{fig:single_kernel_test} bottom panel). We find that the number of false positives is largely unaffected by variations in the gaussian kernel width, except at very low values (Figure~\ref{fig:single_kernel_test} bottom right panel). However the number of false negatives depend on the width of the kernel, rising sharply for values higher than $\sim$3.8\,\AA\ and for values lower than $\sim$2.3\,\AA (Figure~\ref{fig:single_kernel_test} bottom left panel). This is expected, given that MUSE is unable to resolve spectral peaks separated by less than approximately 2.4\,\AA\ due to the limitations of spectral resolution of the instrument. Therefore, at separations below 2.4\,\AA\, double peaks are frequently misidentified as single peaks, leading to an increased false negative rate for double-peaked features. When the kernel width becomes much larger (greater than 3.8\,\AA) the correlation becomes weaker and thus the peak will not be detected. Thus false negatives increase with the kernel width.  

We find that the cutoff value $\sim$ 3.7$\sigma$ and the gaussian width $\sim$ 2.36\,\AA\ represent the values with the best compromise between low false positives and low false negatives and these are the parameters adopted in this study.

\begin{figure}
    \centering
    \includegraphics[width=0.98\linewidth]{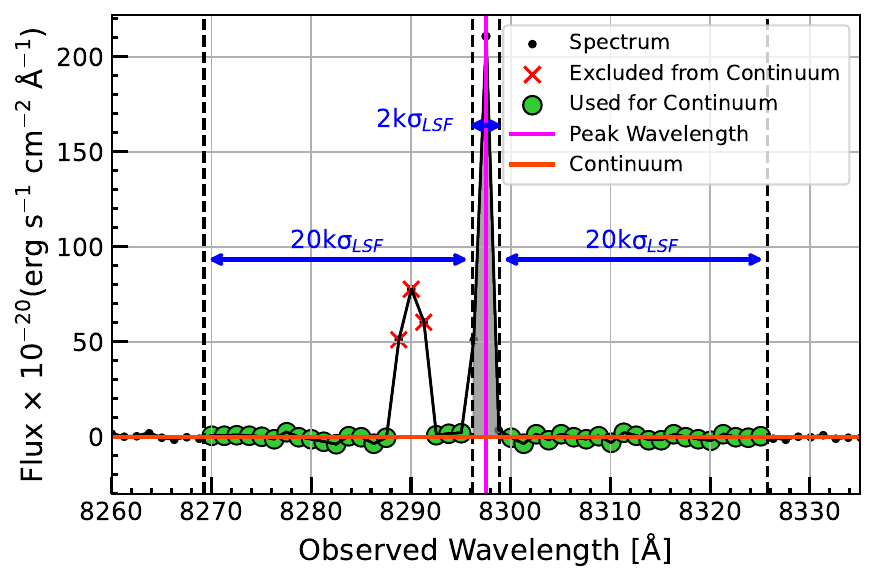}
    \caption{Illustration of the flux estimation of an emission line, with the peak wavelength indicated in magenta. The continuum is estimated using the green points, which lie below 3$\times$ the standard deviation. Red points are excluded from the continuum fit due to significant deviations. The blue arrow away from line indicates the wavelength range on either side of the emission line used to estimate the continuum based on the median flux. The grey shaded region marks the integrated flux of the line.}
    \label{fig:Flux_line}
\end{figure}

\subsection{Estimating limiting flux for a spaxel} \label{sec:1sigma_flux}
To reduce the number of spurious or unreliable detections and to define a robust detection threshold, we estimated the limiting flux for point-like sources with unresolved emission lines. This step ensures that only statistically significant peaks are considered for further analysis, thereby enhancing the reliability of the detected sample. The flux is used to compare with the fluxes of the detected double peaks and to filter out false detections. These comparisons are discussed in detail in Section~\ref{sec:detection} and Section~\ref{sec:binning}.

We estimated the limiting flux for a spaxel for an unresolved emission line similar to \citet{Bacon2023A&A...670A...4B} using the formula :

\begin{equation}
F_{pl}(\lambda)= \Delta \lambda \sqrt{ \frac{\pi r^2}{\Delta s^2} V_{line}(\lambda)}
\end{equation}
where $\Delta \lambda$ is the wavelength bin size (1.25\,\AA), r is the 80\% enclosed flux circular radius for the corresponding Moffat PSF model at 7000\,\AA. $\Delta$s is the spaxel bin size (0.2$\arcsec$). We are using single spaxel for the analysis, thus $\frac{\pi r^2}{\Delta s^2}$ $\sim$ 1. V$_{line}$ is defined as follows:

\begin{equation}
    V_{line}(\lambda)=  \sum_{\lambda^\prime =\lambda-k\sigma_{LSF}(\lambda)}^{\lambda+k\sigma_{LSF}(\lambda)} V_{s}(\lambda^\prime ) 
\end{equation}
where V$_{s}(\lambda)$ is the variance of flux in the spaxel at wavelength $\lambda$, the wavelength interval (k = 1.29) was chosen to capture 90\% of the line flux, and $\sigma_{LSF}(\lambda)$ is derived from the line spread function (LSF) as follows:

\begin{equation}
    \sigma_{LSF}(\lambda) = \frac{\textrm{LSF}(\lambda)}{2\sqrt{\textrm{2ln(2)}}}
\end{equation}
where LSF($\lambda$) is estimated as follows:

\begin{equation}
    \textrm{LSF}(\lambda)= 5.866\times10^{-8}\lambda^2 - 9.187\times10^{-4}\lambda+6.040
\end{equation}
The LSF and $\lambda$ are in \,\AA. The LSF gives the instrumental broadening of the spectral lines.

We also estimated the fluxes of the detected double peaks to compare their fluxes values with F$_{pl}$. We measured the integrated flux of each emission line using the same window interval applied for the F$_{pl}$ estimation of point-like sources. Figure~\ref{fig:Flux_line} shows an example of flux estimation for detected emission line peaks. To determine the local continuum, we selected regions well outside the line profile, between k$\times\sigma_{LSF}$ and 21$\times$k$\sigma_{LSF}$ from the line center on both sides (shown in blue arrow). We estimated the continuum level using 1 $\sigma$-clipped statistics (green points) to robustly exclude outliers (red crosses). After subtracting the continuum (shown in orange), we measured the line flux by integrating the continuum-subtracted spectrum over the defined line region (grey shaded area) using the trapezoidal rule. This approach provides a reliable flux measurement while minimizing contamination.

\section{Application to MUSE-MXDF data and null result}\label{sec:application}
\subsection{Automated Double peak detction Algorithm on MUSE-MXDF Data} \label{sec:detection}
We used MUSE-MXDF dataset (explained in section~\ref{sec:data}) to detect double peak H$\alpha$ signatures. The MUSE-MXDF offers exceptionally deep observations with high S/N ratios, making it an ideal dataset for detecting faint and reliable double-peaked emission features. The MUSE MXDF data exhibit significant spectral masking below 6500\,\AA\, while the region beyond 8950\,\AA\ is largely dominated by noise. Since our goal is to detect double-peaked emission features associated with H$\alpha$, we excluded data below 6500\,\AA. As a result, our study is restricted to the wavelength range 6500–8950\,\AA.

\begin{figure}
    \centering
    \includegraphics[width=0.98\linewidth]{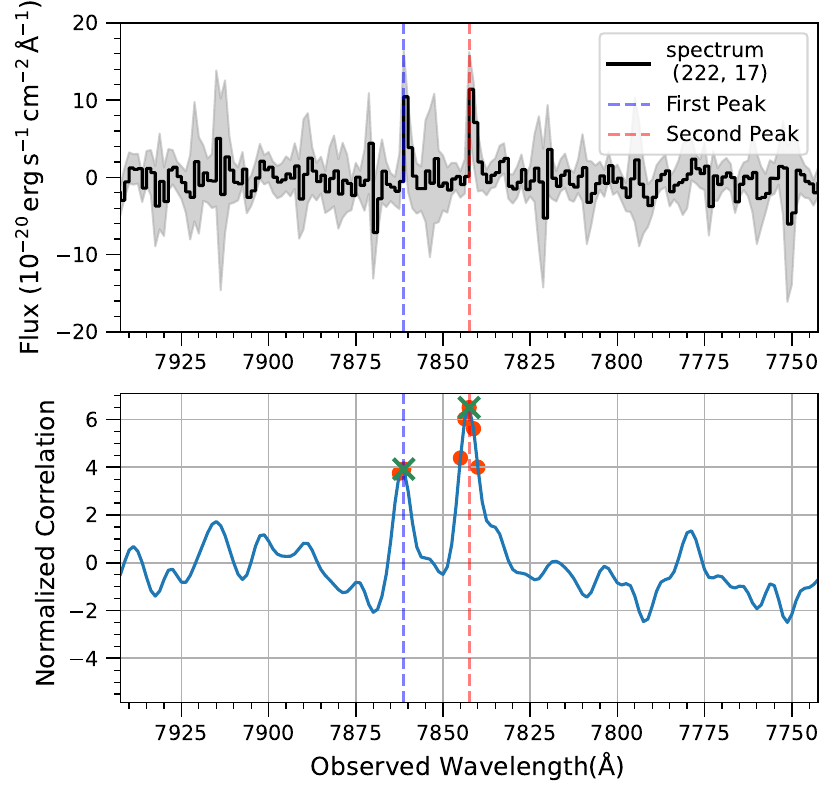}
    \caption{Example of a detected double peak in the observed spectrum (top panel) and the corresponding correlation spectrum (bottom panel). Data points exceeding the 3.7$\sigma$ threshold are shown in orange, while the detected double peaks in the correlation are marked with green crosses. Grey shaded region in the top panel shows the standard deviation of the flux.}
    \label{fig:dp_in_spectrum_correlation}
\end{figure}

\begin{figure}
    \centering
    \includegraphics[width=0.95\linewidth]{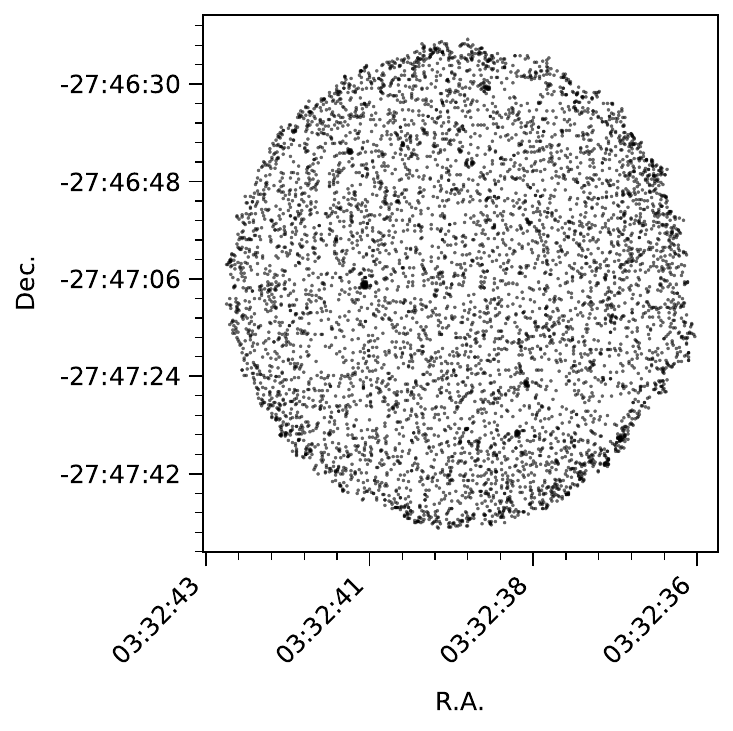}
    \caption{The distribution of detected double peaks in MUSE-MXDF.}
    \label{fig:spatial_distribution_dp}
\end{figure}

\begin{figure*}
    \centering
    \includegraphics[width=0.98\linewidth]{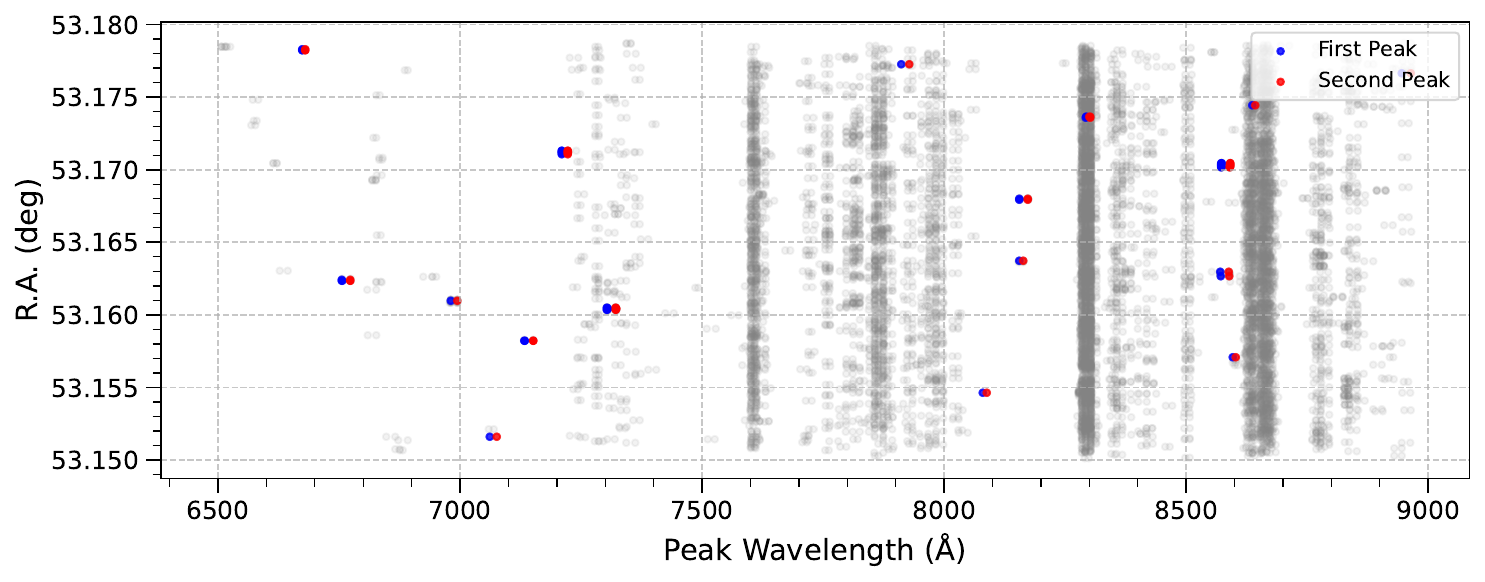}
    \caption{Wavelength versus R.A. for double peaks above the 3F$_{pl}$ threshold. Blue and red represents the first and second peak. Grey shows all the detected peaks.}
    \label{fig:3sigma_wavelength_x}
\end{figure*}

\begin{figure*}
    \centering
    \includegraphics[width=0.98\linewidth]{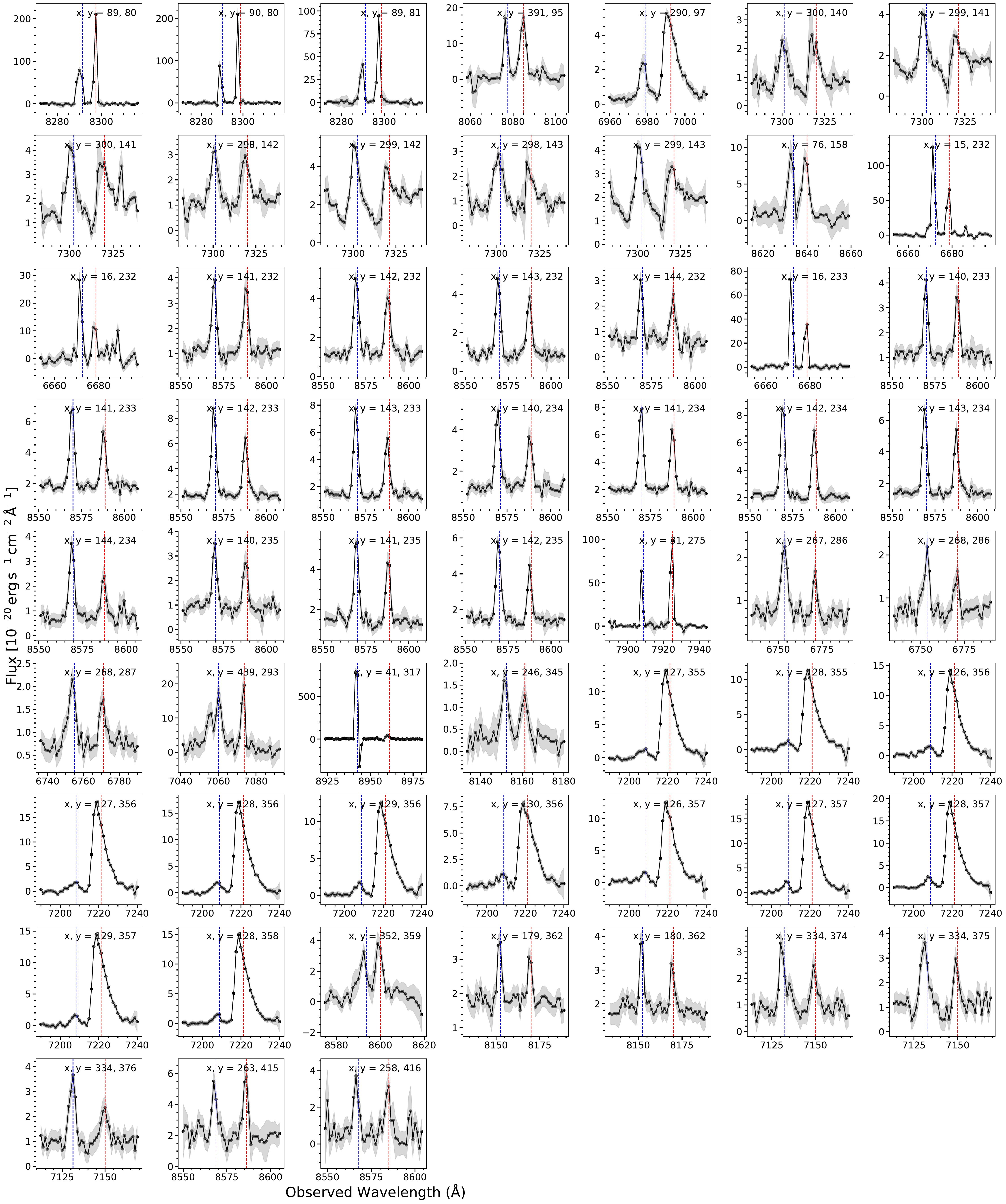}
    \caption{Double peak detections above 3F$_{pl}$ flux threshold in MUSE MXDF from 59 spaxels. Blue and red dashed lines shows the detected double peak locations. The x and y axes in each subplot indicate the spaxel numbers. The grey-shaded region represents the standard deviation of the flux within each spaxel.}
    \label{fig:dp_above_3sigma}
\end{figure*}

\begin{figure*}
    \centering
    \includegraphics[width=0.95\linewidth]{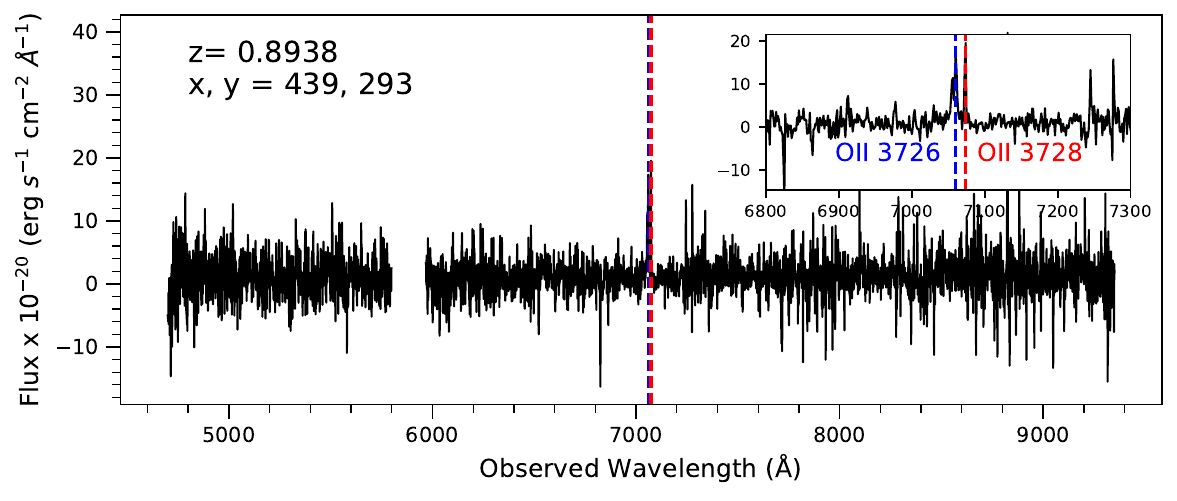}
    \caption{Spectrum of one of cross-matched source. The detected double peaks correspond to the [OII] doublet $\lambda\lambda$3726, 3729 at z = 0.8938. The blue and red dashed lines shows the detected double peaks. The inset shows a zoomed-in view of the spectrum over the wavelength range 6800–7300\,\AA. The x and y are the spaxel coordinates.}
    \label{fig:spectra_crossmatched_sources}
\end{figure*}

\begin{figure*}
    \centering
    \includegraphics[width=0.485\linewidth]{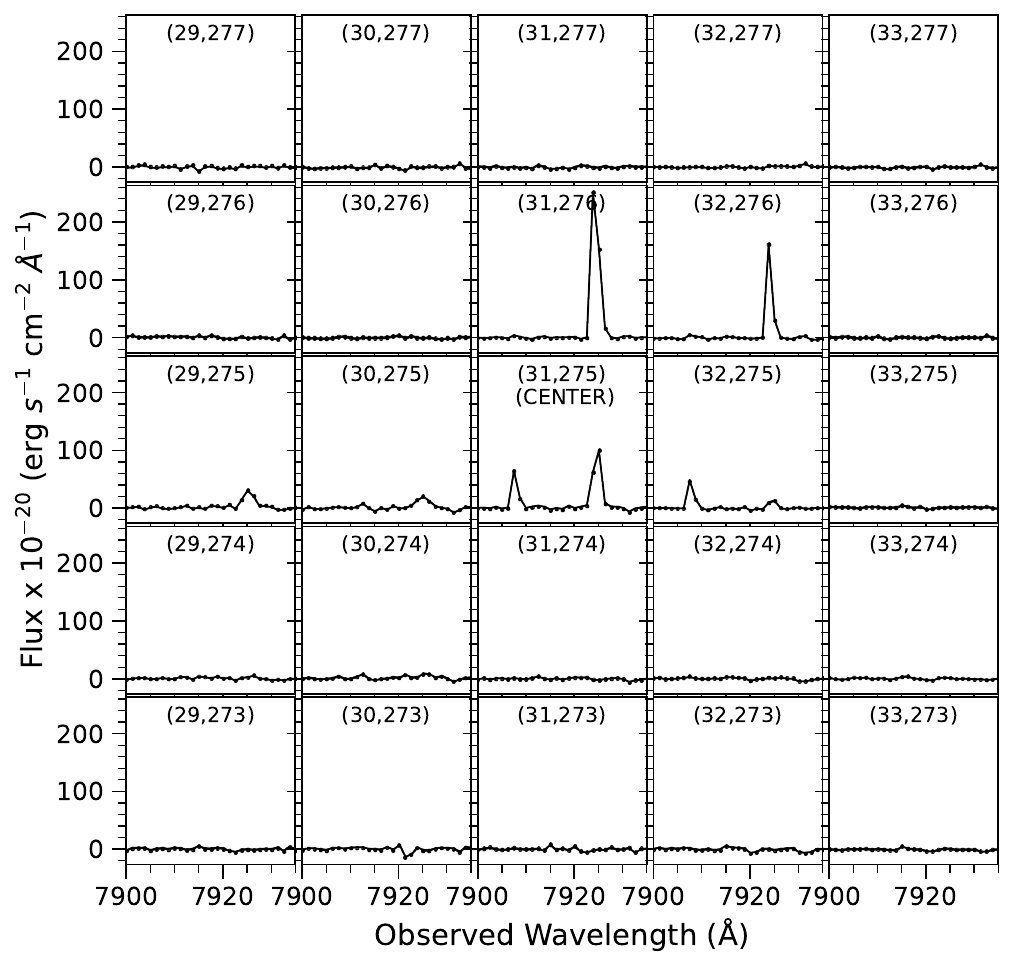}
    \includegraphics[width=0.47\linewidth]{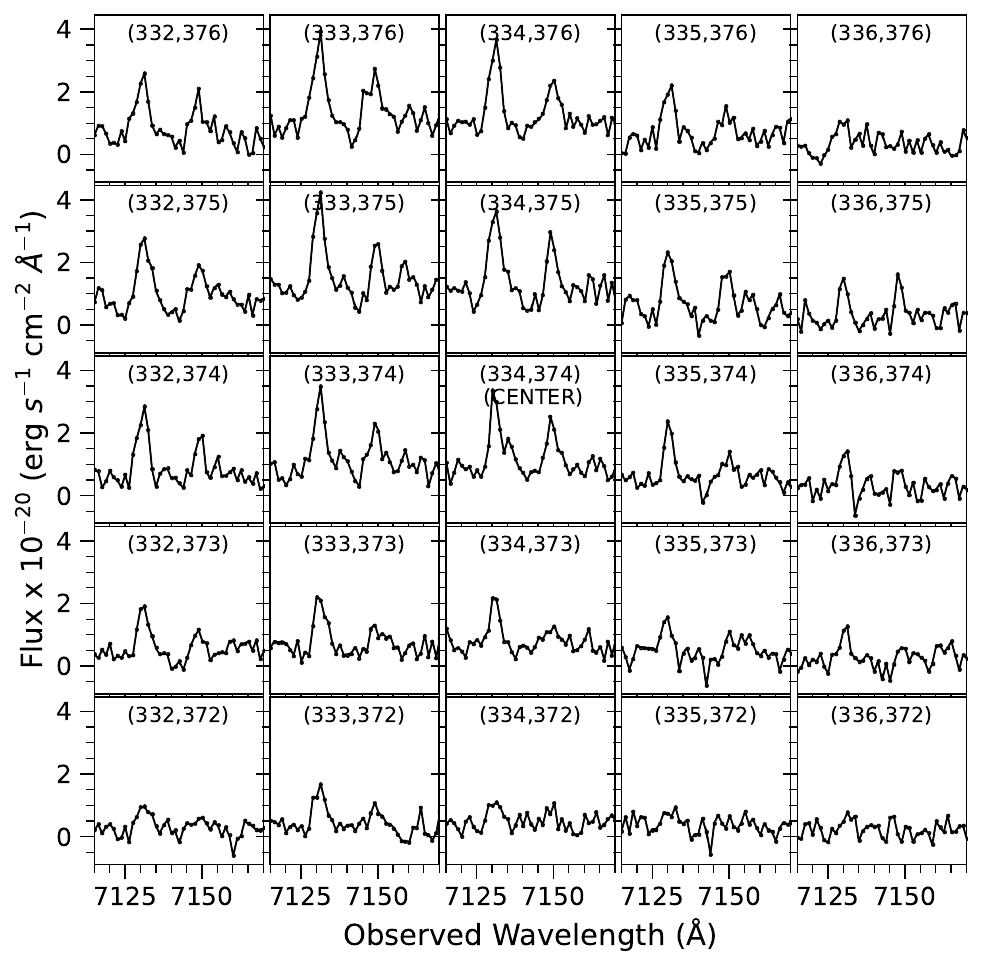}
    \caption{Spectra from neighbouring spaxels surrounding the double-peak detection. The numbers in each subplot are the spaxel coordinates. Each spaxel is separated by 0.2\arcsec, so that all 25 spaxels fall within a single spatial resolution element and are expected to exhibit similar properties if they originate from the same astronomical source. The left and right panels show the spectra from neighbouring spaxels for an unmatched and a matched double-peak detection, respectively.} 
    \label{fig:cosmic_ray}
\end{figure*}

We applied our automated double-peak detection algorithm to the MXDF data to systematically identify and analyze double peak features from each spaxel. An initial search yielded 6094 double-peak detections, including some spaxels exhibiting multiple double peaks. Figure~\ref{fig:dp_in_spectrum_correlation} shows one such example of a detected double peak in the observed spectrum (top panel) and the corresponding correlation spectrum (bottom panel). Figure~\ref{fig:spatial_distribution_dp} shows the spatial distribution of the detected double peaks.
This unexpectedly high detection rate suggests the presence of unreliable or spurious detections. We expect the sources to span a continuous range of redshifts, and thus double peaks should appear at varying observed wavelengths. To investigate the reliability of the detections, we examined the wavelength distribution of the double peaks as a function of the Right Ascension (R.A.) of the spaxel (Figure~\ref{fig:3sigma_wavelength_x}). We found a strong clustering of detections at specific wavelengths (grey points), indicating that many of these are unlikely to be real astrophysical sources. To mitigate false detections, we applied an additional selection criterion requiring that both peaks in each candidate must individually exceed the 3F$_{pl}$ threshold (explained in Section~\ref{sec:1sigma_flux}) for point sources. We considered a detection reliable only if both peaks exceeded the 3F$_{pl}$ threshold for a point-like source. After the cut, we obtained 59 spaxels with reliable double-peak detections. The blue and red points shows the wavelengths of the first and second detected peaks above the $3F_{\rm pl}$ flux threshold (Figure~\ref{fig:3sigma_wavelength_x}). Figure~\ref{fig:dp_above_3sigma} shows the spectra at the locations of these double-peak detections, highlighting the detected features.

To further investigate the origin of the detected double peaks having flux above 3F$_{pl}$, we cross-matched their positions with sources from the catalogue provided by \citet{Bacon2023A&A...670A...4B} for the MUSE-MXDF field. We adopted a matching radius of 1$\arcsec$, which is appropriate given the spatial resolution of the MUSE observations and the typical astrometric uncertainties. We find that 50 out of 59 spaxels correspond to 13 sources in the catalogue. 9 spaxels do not correspond to any source. The full spectrum of one of the crossmatched sources is shown in Figure~\ref{fig:spectra_crossmatched_sources} as an example. The detected double peaks correspond to the [OII] $\lambda\lambda$3726, 3729 doublet emitted by a galaxy at a redshift of $z=0.8938$. Of the 13 sources exhibiting apparent double-peaked features, most can be explained by known emission-line doublets: three [\ion{S}{ii}], three [\ion{Mg}{ii}], two Ly$\alpha$, three [\ion{O}{ii}], and one [\ion{Ca}{ii}]. The last of the 13 sources exhibits a double-peaked feature at wavelengths consistent with the expected H$\alpha$ emission of the cross-matched source. However, a detailed spatial-spectral analysis indicates that, one H$\alpha$ component is PSF-like and spatially extended, whereas the second  peak is confined to only two spaxels, consistent with a cosmic-ray artifact. This shows that, while some double-peak detections are likely artefacts or noise fluctuations, the double peaks above 3F$_{pl}$ threshold are genuine spectral features associated with known sources, validating the effectiveness of our cross-matching approach in identifying reliable astrophysical candidates.

We further examined these nine unmatched double-peak detections that were not associated with any known sources, along with their neighbouring spaxels. The spectra of these unmatched spaxels show sharp, isolated peaks at the same wavelengths as the detected double peaks (Figure\ref{fig:cosmic_ray}, left panel). For comparison, we examined spaxels associated with catalogued sources (Figure \ref{fig:cosmic_ray}, right panel). In the crossmatched cases, the emission peaks are spatially extended and consistent across neighboring spaxels, following the point-spread function. In contrast, peaks detected in unmatched spaxels exhibit significantly higher fluxes but are confined to only a few spaxels, suggesting a likely instrumental or cosmic-ray origin. These characteristics may indicate cosmic ray contamination or something equivalent. Similar features were identified in other unmatched cases, further supporting the interpretation that these detections are likely due to cosmic ray hits.

Additionally, many spaxels with double peaks below the 3$\sigma$ flux threshold likely result from photon noise, residual sky subtraction errors, telluric contamination, or systematic instrumental effects, all of which can produce spurious features in the dataset. Unfortunately, we do not identify any new sources that could be considered potential IMBH candidates.

\subsection{Binning the Data}\label{sec:binning}
To further seek for possible IMBH candidates we enhanced the S/N by binning the MUSE-MXDF data. MUSE-MXDF has spatial resolution of approximately 0.7$\arcsec$ and a pixel scale of 0.2$\arcsec$, we applied a 4$\times$4 spatial binning before rerunning the automated double-peak detection algorithm. We identified 545 double-peaks. Applying a 1F$_{pl}$ point-source flux threshold reduced the sample to 28 detections. We cross-matched the detections with the MUSE-MXDF source catalogue using a 1$\arcsec$ matching radius and found that 23 spaxels matched with sources in \citep{Bacon2023A&A...670A...4B} catalogue. As multiple spaxels can be associated with a single catalogue source, these correspond to a total of 11 objects. Among them, four are identified as [\ion{S}{ii}] doublets, two as Ly$\alpha$, two as [\ion{O}{ii}] doublets, and two as [\ion{Mg}{ii}] doublets. One source does not exhibit double-peaked features at any known redshifted emission line of the matched object. A detailed inspection of the spectra shows that the apparent double peaks in this case are confined to only two spaxels, identical to those affected by cosmic-ray artifacts in the unbinned data. This suggests a non-astrophysical origin, and therefore none of the binned detections are considered viable IMBH candidates. Among the five unmatched spaxels, one was associated with two catalogue sources, however the detected double peak did not correspond to the redshifted wavelength of either source. A closer inspection indicates that this feature is consistent with a cosmic-ray artifact, as also seen in the unbinned data. The remaining four unmatched spaxels lie near the edge of the field, where increased noise levels may contribute to spurious detections. Thus, we did not identify any new double-peak features that could be attributed to IMBH candidates in the binned MUSE-MXDF data. As a result, the analysis yielded a null detection of new IMBH candidates in the binned dataset.

\begin{figure}
    \centering
    \includegraphics[width=0.98\linewidth]{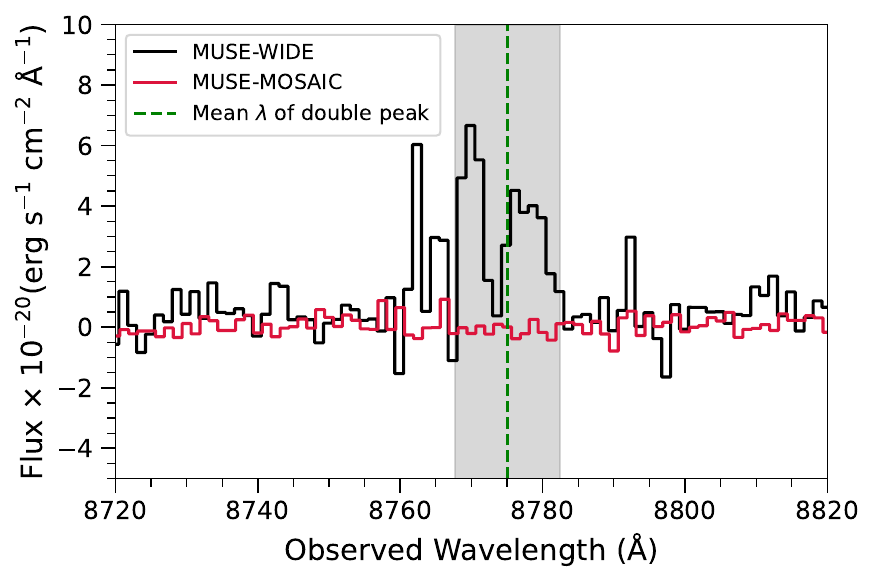}
    \caption{Spectrum of one of the candidate IMBH from \citet{Almeida2022ApJ...934..100S} in the new MUSE-MOSAIC data (red) and the original MUSE-WIDE data (black).}
    \label{fig:overlapping_candidate}
\end{figure}

\begin{figure*}
    \centering
    \includegraphics[width=0.98\linewidth]{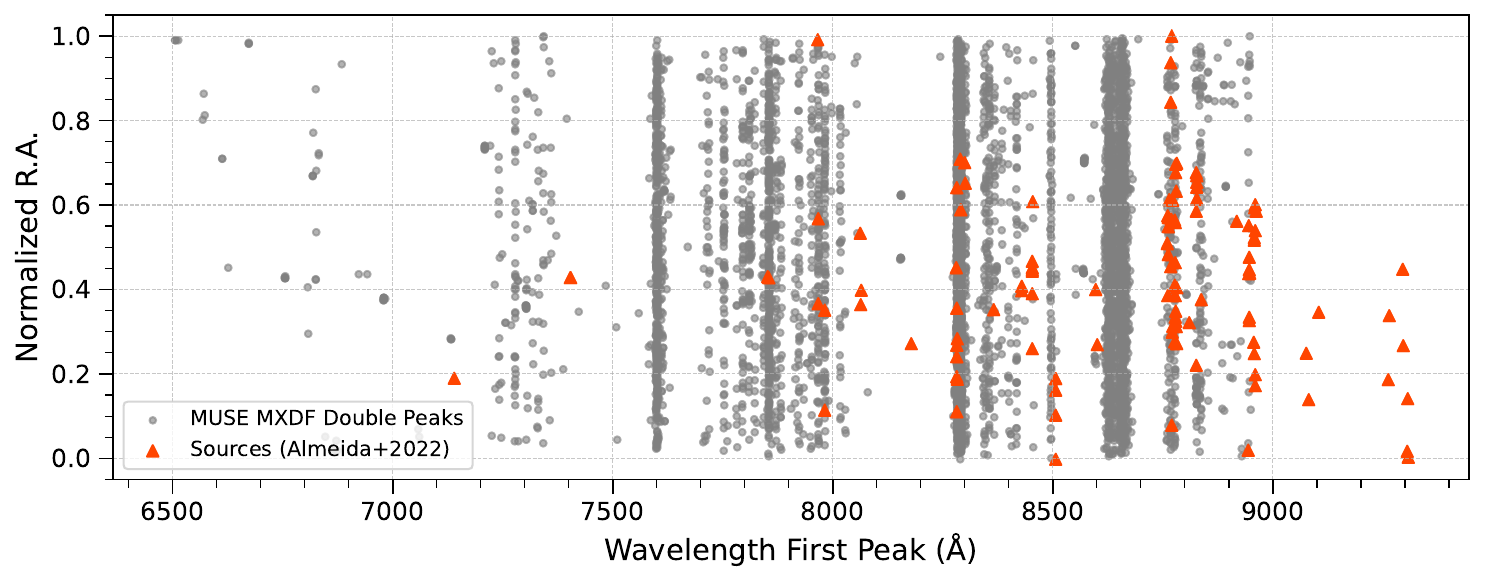}
    \caption{Comparison of double peak from \citet{Almeida2022ApJ...934..100S} (red) with the detected double peaks from MUDF (grey). MUSE-WIDE covers a larger field of view which is spatially separated from MUSE-MXDF, so we normalize the R.A. range to visualize the dataset together.}
    \label{fig:comparison_almeida}
\end{figure*}

\section{Discussion}\label{sec:discussion}

\subsection{Reassessing Previously Reported IMBH Candidates from \citet{Almeida2022ApJ...934..100S}}\label{sec:comparison}. We checked whether the IMBH candidates reported by \citet{Almeida2022ApJ...934..100S} in the MUSE-Wide survey are also detected in the MUSE-MOSAIC field, which partially overlaps with MUSE-Wide. Out of the 118 candidates identified in MUSE-Wide, five lie within the overlapping region. We extracted spectra at the positions of these five sources from MUSE-MOSAIC using the same method as in the MUSE-Wide analysis. However, none of the five spectra from MUSE-MOSAIC exhibited double-peaked H$\alpha$ features (Figure~\ref{fig:overlapping_candidate}). To ensure that this discrepancy is not due to astrometric misalignment or extraction errors, we verified the astrometric accuracy between the two fields by comparing the spectra of a bright galaxy present in both datasets. The spectra from MUSE-Wide and MUSE-MOSAIC for this galaxy show agreement, confirming consistent wavelength and spatial alignment between the fields.

We also compared the wavelength of the sources identified in MUSE-Wide \citep{Almeida2022ApJ...934..100S} with those detected in the MUSE-MXDF. The candidates in MUSE-Wide also exhibit an overdensity at the same wavelengths as observed in the MUSE-MXDF field (Figure~\ref{fig:comparison_almeida}). Most of these sources ($\approx$ 85\%) align with the noisy spectral regions identified in the MUSE-MXDF data, suggesting that some of the double-peak detections in \citealt{Almeida2022ApJ...934..100S} could be due to noise/artifacts in the data which is why we could not recover them in MUSE-MOSAIC data. We discuss potential sources of artifacts in Sec.~\ref{sec:artifacts}.

\subsection{Sky Subtraction Issues} \label{sec:artifacts}
The reason for the detection of such a large number of double peaks (6094) in the MUSE-MXDF data could be related to sky subtraction issues inherent to the pipeline-processed data. The bright sky emission lines often leave significant residuals after the standard sky subtraction, which can interfere with the detection of faint emission features. To mitigate these residuals, several specialized post-processing tools such as ZAP \citep{Soto2016MNRAS.458.3210S}, CubEx \citep{Cantalupo2019MNRAS.483.5188C}, and CubePCA \citep{Husemann2022A&A...659A.124H} are commonly employed by the MUSE community. These packages are specifically designed to improve the removal of telluric emission lines, especially in deep observations targeting faint emission-line sources in largely empty sky regions.

However, while these additional sky-subtraction methods can significantly reduce telluric residuals, they may also introduce unintended alterations to the data. For example, it has been reported that ZAP can, in some cases, partially remove the continuum of a source or modify intrinsic spectral features \citep{Weilbacher2022SPIE12189E..12W}. Such alterations may potentially suppress or distort the double-peaked signatures we aim to detect.

\subsubsection{Flat Field Issues}
Another issue in MUSE data is the presence of spatial patterns in the images generated by integrating over multiple wavelength planes from the reduced datacubes. These patterns, caused by the 24 individual slicer integral field units in MUSE, appear as alternating bright and dark regions that vary with wavelength \citep{Weilbacher2022SPIE12189E..12W}. Additional small-scale artifacts arise from the arrangement of slicer stacks and their boundaries, where isolated pixels can appear unusually bright or dark \citep{Weilbacher2022SPIE12189E..12W}.

Although these spatial artifacts and the sky subtraction residuals differ in nature, both are closely tied to flat-fielding accuracy. 
Several calibrations, including the line-spread function, wavelength calibration, OH-transition flux ratios, and flux homogeneity, were tested for their impact on sky subtraction accuracy. Field-wide flux variations of ~1.7\% were found \citep{Weilbacher2022SPIE12189E..12W}, comparable to the variations in the sky background.
These artifacts can significantly affect the detection of faint or complex features, such as double peaks, especially in automated searches. Addressing these instrumental effects is critical for reliable data interpretation.

\section{Conclusions}\label{sec:conclusion}
In this work, our aim was to identify the spectroscopic signatures of accreting, wandering IMBHs through double-peaked H$\alpha$ emission up to a redshift of $\sim$0.35. To achieve this, we developed an automated detection algorithm that cross-correlates a single Gaussian kernel with the observed spectra to identify double-peaked features (Section~\ref{sec:simple_kernel}), and applied it to the MUSE-MXDF dataset to search for new potential IMBH candidates (Section~\ref{sec:application}). To maximize the performance of the algorithm, we systematically tested the impact of varying the gaussian width and detection cutoff, selecting the parameters that minimized the false detection rate (Section~\ref{sec:effectiveness_simple_kernel}). Based on these tests, we adopted a cutoff threshold of 3.7$\sigma$ and a gaussian width of 2.4\,\AA\ for the final search.
To ensure the reliability of the detections, we applied an additional 3F$_{pl}$ flux threshold for point-like sources, retaining only those double peaks where both components exceeded this flux limit (Figure~\ref{fig:dp_above_3sigma}). After careful inspection and cross-matching with existing source catalogues, we identified the spaxels with double-peak detections that did not correspond to any known sources. Further analysis of these cases indicated that the detected features are most likely caused by cosmic ray contamination rather than true astrophysical signals (Figure~\ref{fig:cosmic_ray}). Despite the exceptional depth and high S/N ratio of the MXDF dataset, our automated search did not yield any new reliable double-peak detections that could be associated with IMBH candidates. To further improve sensitivity, we applied 4$\times$4 spatial binning to enhance the S/N, however, this approach did not yield any additional IMBH candidates (Section~\ref{sec:binning}). We found that all detections were either associated with previously known sources in the catalogue provided by \citet{Bacon2023A&A...670A...4B}, or attributable to cosmic ray hits, residual sky background, flat-fielding artifacts, or edge-of-field noise contamination. After carefully examining the spectral and spatial characteristics of these detections, we find no compelling evidence for new, isolated IMBH candidates in our dataset. Therefore, we report a null result for the identification of IMBHs via double-peaked emission-line profiles.

While no new candidates were recovered, the outcome remains scientifically informative. First, it underscores the inherent complexity and observational challenges involved in detecting wandering IMBHs, particularly in deep-field datasets where such sources are expected to be faint and rare. Second, it highlights the necessity of rigorous validation procedures to distinguish genuine astrophysical signals from artifacts in integral field spectroscopy data. Third, our results demonstrate the practical utility and robustness of automated detection algorithms for systematic searches. Even though our approach did not yield positive detections in this case, it offers a scalable and objective framework that can be applied to larger or complementary datasets e.g, Dark Energy Spectroscopic Instrument survey, James Webb Space Telescope and future facilities such as 4MOST. As observational capabilities and data volumes continue to grow, such automated methodologies will become increasingly valuable in identifying elusive IMBH populations and constraining their abundance and distribution in various galactic environments.

\begin{acknowledgments}
We thank the anonymous referee for their insightful review, which has substantially enhanced the clarity and impact of this work. The authors acknowledges financial support from the Spanish Ministry of Science, Innovation and Universities (Ministerio de Ciencia, Innovación y Universidades, MICIU) , project PID2022-136598NBC31 (ESTALLIDOS8) by MCIN/AEI/10.13039/501100011033. It was also supported by the European Union through the grant “UNDARK” of the widening participation and spreading excellence programme (project number 101159929). This paper has used the observations collected at the European Southern Observatory under ESO programme 1101.A-0127, 094.A-0289, 095.A-0010, 096.A-0045, 094.A-0205, 095.A-0240, 096.A-0090, 097.A-0160 and 098.A-0017. The authors thank Tanya Urrutia for clarifications regarding the MUSE data cubes and Peter
Weilbacher and Roland Bacon for the data reduction. 
\end{acknowledgments}




%
\facilities{MUSE \citep{Bacon2014Msngr.157...13B}}

\software{astropy \citep{AstropyCollaboration2013A&A...558A..33A}, NumPy \citep{Harris2020Natur.585..357H}, MPDAF \citep{Bacon2016ascl.soft11003B}
          }







\bibliography{sample7}{}
\bibliographystyle{aasjournalv7}



\end{document}